\documentclass[]{spie}

\newcommand{\um}{$\mu$m}

\usepackage{amsmath,amsfonts,amssymb}
\usepackage{graphicx}
\usepackage[table, x11names]{xcolor}
\usepackage[colorlinks=true, allcolors=blue]{hyperref}

\title{A novel freeform slicer IFU for the Magellan InfraRed Multi-Object Spectrograph (MIRMOS)}

\author[a]{Maren Cosens}
\author[a]{Nicholas P. Konidaris II}
\author[a]{Gwen C. Rudie}
\author[a]{Andrew B. Newman}
\author[a]{Gerrad Killion}
\author[a]{Leon Aslan}
\author[b]{Robert Barkhouser}
\author[c]{Andrea Bianco}
\author[a]{Christoph Birk}
\author[a]{Julia Brady}
\author[c]{Michele Frangiamore}
\author[a]{Tyson Hare}
\author[d]{Stephen C. Hope}
\author[a]{Daniel D. Kelson}
\author[e]{Alicia Lanz}
\author[a]{Solange Ramirez}
\author[d]{Stephen A. Smee}
\author[c]{Andrea Vanella}
\author[a]{Jason E. Williams}

\affil[a]{Carnegie Science, The Observatories, 813 Santa Barbara Street, Pasadena, CA 91101, USA}
\affil[b]{LCS Optics LLC, Parkton, MD, USA}
\affil[c]{INAF -- Observatorio Astronomico di Brera, via Bianchi 46, 23807, Merate, Italy}
\affil[d]{Johns Hopkins University, Department of Physics and Astronomy, 3701 San Martin Drive,
Baltimore, MD 21218, USA}
\affil[e]{Capella Space, 438 Shotwell St., San Francisco, CA, 94110}

\authorinfo{Further author information: (Send correspondence to M.C.)\\M.C E-mail: mcosens@carnegiescience.edu}
 
\begin{document} 
\maketitle

\begin{abstract}
The Magellan InfraRed Multi-Object Spectrograph (MIRMOS) is a planned next generation multi-object and integral field spectrograph for the 6.5m Magellan telescopes at Las Campanas Observatory in Chile. MIRMOS will perform R$\sim$3700 spectroscopy over a simultaneous wavelength range of 0.886 - 2.404\um\, (Y,J,H,K bands) in addition to imaging over the range of 0.7 - 0.886\um. The integral field mode of operation for MIRMOS will be achieved via an image slicer style integral field unit (IFU) located on a linear stage to facilitate movement into the beam during use or storage while operating in multi-object mode. The IFU will provide a $\rm \sim20''\times26''$ field of view (FoV) made up of $\rm0.84''\times26''$ slices. This will be the largest FoV IFS operating at these wavelengths from either the ground or space, making MIRMOS an ideal instrument for a wide range of science cases including studying the high redshift circumgalactic medium and emission line tracers from ionized and molecular gas in nearby galaxies. In order to achieve the desired image quality and FoV while matching the focal ratio to the multi-object mode, our slicer design makes use of novel freeform surfaces for the pupil mirrors, which require the use of high precision multi-axis diamond milling to manufacture. We present here the optical design and predicted performance of the MIRMOS IFU along with a conceptual design for the opto-mechanical system. 
\end{abstract}


\section{INTRODUCTION} \label{sec:intro} 
The last two decades have seen a massive rise in the use of integral field spectrographs (IFS) for both ground and space based observations. In particular, there has been incredibly high demand for wide-field optical IFSs on large ground based telescopes like Keck/KCWI\cite{Morrissey2018} and VLT/MUSE \cite{Bacon2010}. These instruments have been highly productive across a range of science cases; allowing astronomers to begin efficiently mapping the morphology and kinematics of the circumgalactic medium (CGM), probing stellar feedback in nearby star forming regions, and much more. However, despite this success, an equivalent does not yet exist on large telescopes operating at near infrared (near-IR) wavelengths. This means that despite the groundbreaking observations with these instruments, astronomers cannot yet robustly constrain the source of the CGM emission at high redshift or the impact of different modes of stellar feedback in molecular clouds.

\begin{figure}
    \centering
    \includegraphics[width=0.9\textwidth]{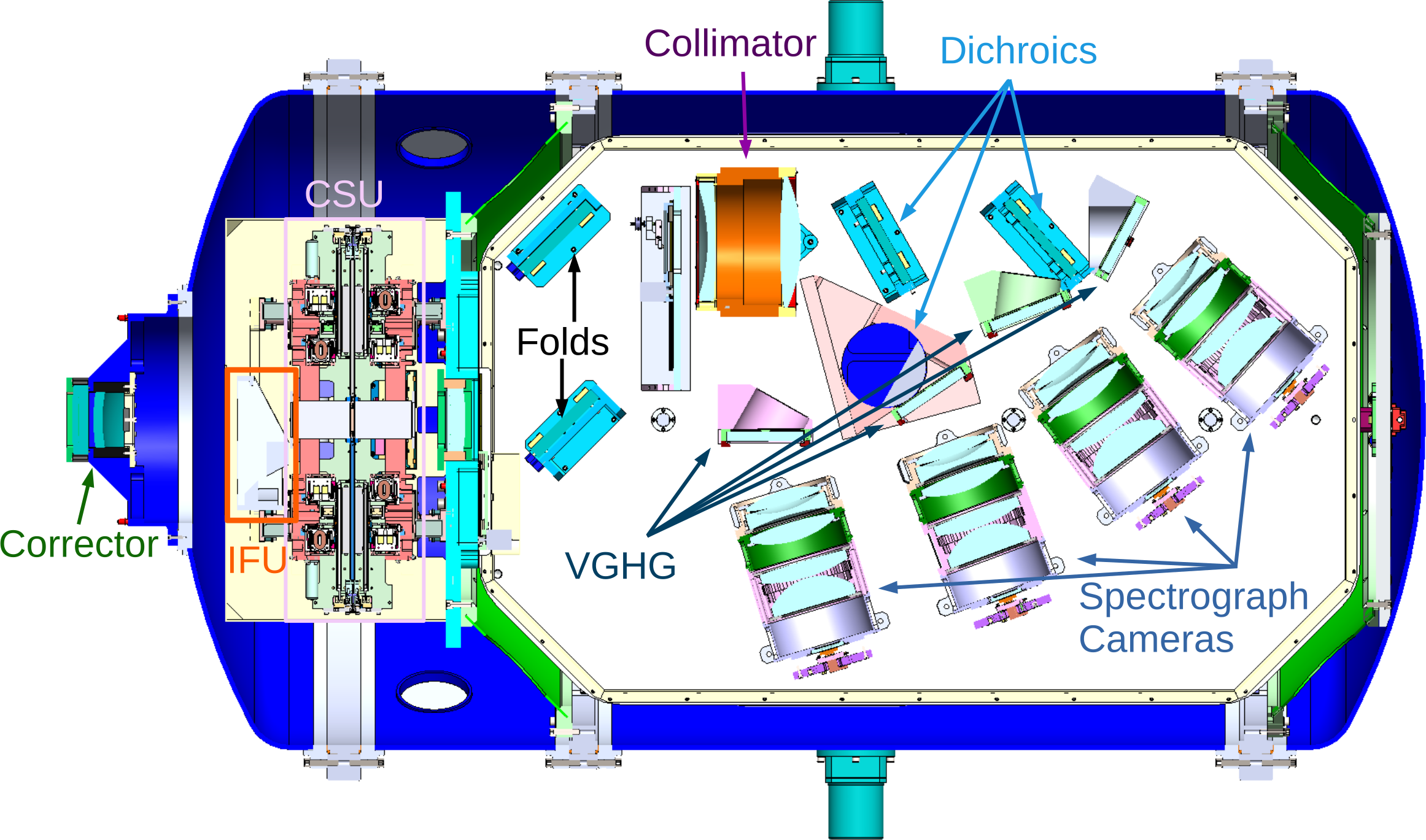}
    \caption{CAD model of the full MIRMOS instrument with key components labeled. Light enters the cryostat from the left through a two-element corrector with the front element doubling as the dewar window. Next, if the IFU is in its active position (shown here stored), light passes through the IFU optics described in Section \ref{sec:optics}. Otherwise, the light passes directly to the slit mask formed by the CSU for either long-slit or MOS mode. The rest of the optics are common between these two modes with the newly formed slits or pseud-slits passing through two fold mirrors and a collimator followed by a dichroic tree to the imaging camera (underneath the optical bench) or the four Y,J,H,K gratings and cameras shown here. Each of these bands has a dedicated detector assembly meaning that all five will be used simultaneously.}
    \label{fig:MIRMOS}
\end{figure}

We are designing a slicer integral field unit (IFU) for the Magellan InfraRed Multi-Object Spectrograph (MIRMOS)\cite{Konidaris2020, Konidaris2022}\footnote{MIRMOS website: https://sites.google.com/carnegiescience.edu/mirmos/overview} to specifically address this need. MIRMOS is a next-generation, near-IR, combination multi-object spectrograph (MOS) and IFS for the $\rm6.5\,m$ Magellan telescopes at Las Campanas Observatory (LCO) in Chile. MIRMOS uses a five-channel design via a series of dichroics in order to perform simultaneous R$\sim$3,700 spectroscopy over a wavelength range of 0.886 - 2.404\um \, (Y, J, H, and K bands) along with imaging from 0.7 - 0.886\um. The MOS mode uses a configurable slit unit (CSU) made up of 92 pairs of masking bars in order to select slit positions and adjust masks in real time over a $\rm 13'\times3'$ field of view (FoV). When the integral field mode is in use, the bars will open fully and the slicer IFU will be moved into the beam as a single monolithic unit. A CAD model of the MIRMOS cryostat highlighting the location of the IFU is shown in Figure \ref{fig:MIRMOS}. After passing through either the IFU or CSU, light enters the collimator and the dichroic tree which feeds the five operating channels: a modified i-band through slit imager (0.7 - 0.886\um), and the Y (0.886 - 1.124\um), J (1.124 - 1.352\um), H (1.466 - 1.807\um), and K (1.921 - 2.404\um) spectrograph channels. Each spectrograph channel will have an individual, high-throughput, volume phase holographic (VPH) grating, a fast $\rm f/1.4$ camera, and a dedicated Teledyne Hawaii-2RG detector. For more details of the MIRMOS design see Konidaris et al. (2020\cite{Konidaris2020}, 2022\cite{Konidaris2022}, and \textit{this conference}\cite{Konidaris2024}).

The MIRMOS IFU is specifically designed to address the aforementioned lack of wide-field IFSs operating at near-IR wavelengths. The design consists of a slicer style IFU with twenty-three, $\rm0.84''\times26''$ slices for a total FoV of $\rm19.32''\times26''$. In order to achieve the desired image quality and FoV while respecting the mechanical and optical constraints set by the MOS mode, this IFU design makes use of freeform optics. The key science cases driving these requirements are discussed in Section \ref{sec:science}. The optical design of the IFU is described in Section \ref{sec:optics}, and a conceptual mechanical design is shown in Section \ref{sec:mech}.

\section{SCIENCE DRIVERS} \label{sec:science}
MIRMOS --- and its IFU --- will be a facility instrument at Magellan and is therefore designed for use across a wide range of science cases. Still, the requirements for the IFS mode are primarily driven by two key science cases: (i) robustly measuring the morphology and kinematics of the $\rm 1<z<3$ CGM, and (ii) probing stellar feedback in the star forming regions of nearby galaxies.

Both of these science cases above --- and others such as studies of globular clusters, or observing multiple lensed galaxies --- benefit from a large FoV akin to that of the optical IFSs like KCWI\cite{Morrissey2018}, MUSE\cite{Bacon2010}, and LLAMAS\cite{Furesz2020}. However, another critical goal is to maximize the spatial and spectral resolution in order to disentangle the motion of distinct gas clumps in both the CGM and star forming regions. The slice width chosen is the smallest that can be fully sampled at the detector at $\rm 0.84''$, and provides a strong balance between maximizing the resolution and FoV. This has an added benefit of matching the nominal slit width to be used in the MOS mode, giving a consistent spectral resolution between these two primary instrument modes.

\subsection{Requirements} \label{sec:req}

The main requirements for the MIRMOS IFU are listed in Table \ref{tab:IFU_reqs} with their driving science case and design impact. Some of these requirements are also driven by the practical constraint of fitting into the instrument around space required for the CSU on one side and the front bell of the instrument on the other. This drives the maximum volume and the minimum distance between the optics/mechanism and the focal surface; this will be discussed further in Section \ref{sec:optics}.

\begin{table}[h]
\centering
\caption{Key requirements for the MIRMOS integral field mode} 
\label{tab:IFU_reqs} 
\includegraphics{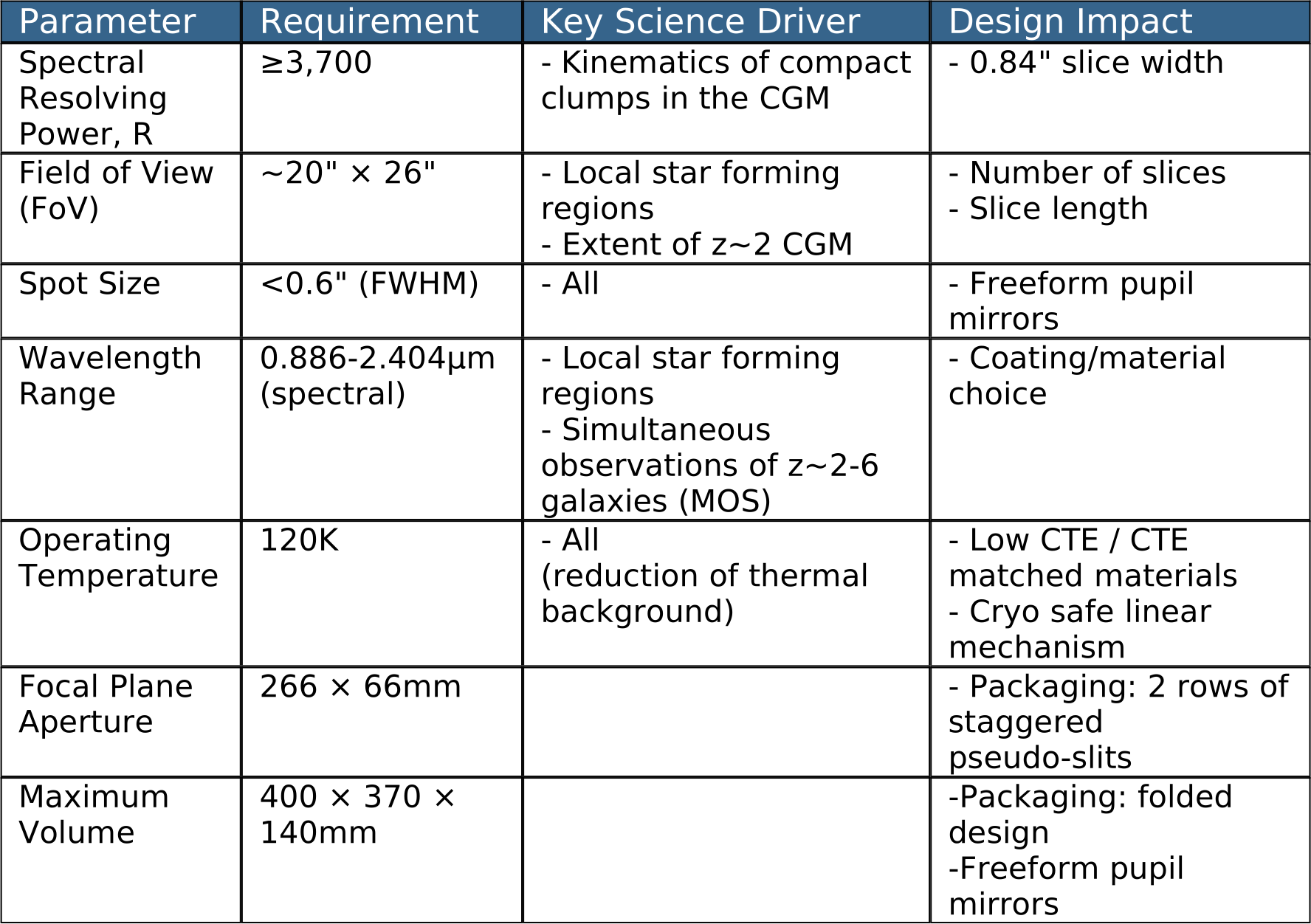}
\end{table}

\section{OPTICAL DESIGN} \label{sec:optics}
\begin{figure}
    \centering
    \includegraphics[width=0.7\textwidth]{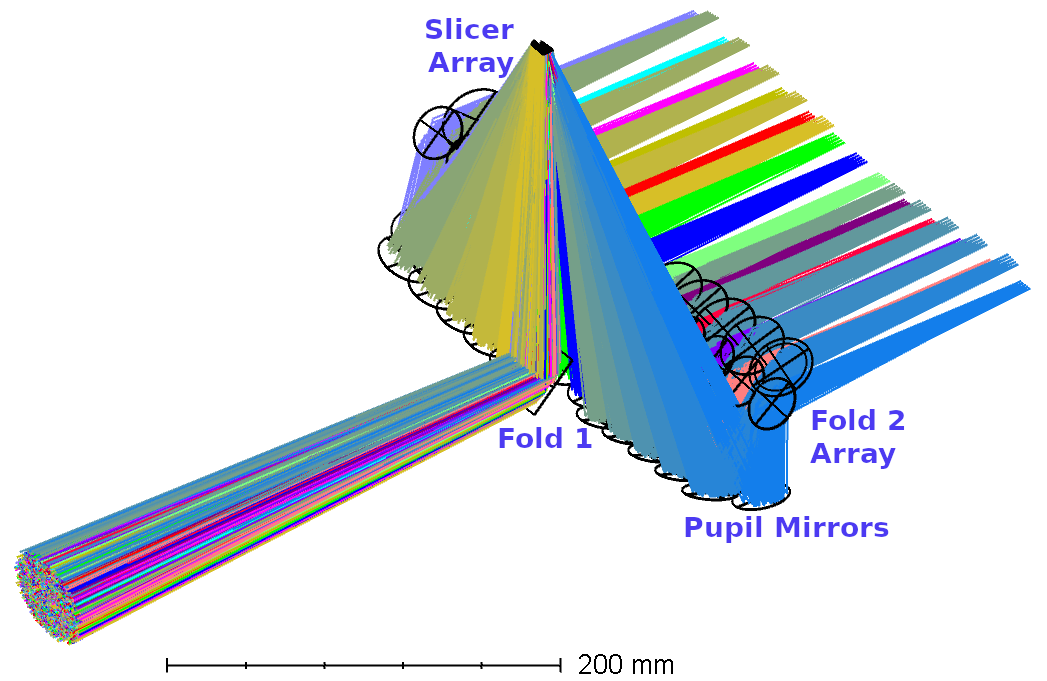}
    \caption{Zemax model + raytrace of the full 23 slice IFU system. From the left, the incoming beam is directed up at a 90$^\circ$ angle by a flat fold mirror where the field is divided up into 23 segments (or “slices”) by the array of slicer mirrors. The slicer mirrors then direct each segment (represented by the unique colors) back down to the pupil mirrors at increasing angles in order to fan out into 2 rows of distinct pupil mirrors. These pupil mirrors direct each slice up to a flat fold mirror angled such that the beam after the fold is parallel to the incoming beam. The focus of each slice is at the same corrected focal surface of the MOS mode. This forms two rows of staggered pseudo-slits which feed into the rest of the spectrograph.}
    \label{fig:IFU_layout}
\end{figure}

The MIRMOS IFU design utilizes a slicer style with all reflecting optics. There are four optical components of the IFU: (i) a flat fold mirror which allows enough space for the system before the CSU mechanism; (ii) a monolithic slicer array of 23 individual segments; (iii) an array of 23 freeform pupil mirrors; and (iv) an array of 23 flat fold mirrors to re-align the beam. The IFU field is located $\rm 1.3'$ off-axis. This is necessary in order to create sufficient space between elements (iii) and (iv) while keeping the beam as close to the center of the CSU aperture as possible. A Zemax raytrace of the optical layout is shown in Figure \ref{fig:IFU_layout}.

The IFU performs 1:1 reimaging at an f/11 focal ratio. Each of the 23 ``slices" has a unique optical prescription for both the slicer and pupil mirror, with freeform surfaces on the pupil mirrors. The slicer mirrors are each $\rm 0.84''\times26''$ ($\rm\sim0.35mm\times9mm$) parabolic or hyperbolic mirrors with radii of curvature ranging from 330 - 545 mm. The choice of slice width allows us to maximize the spectral resolution as needed for the key MIRMOS science cases while still fully sampling the PSF. The $\rm 0.84''$ slices correspond to two pixels on the spectrograph detectors; sampling will be increased using the integrated dithering stage (Smee et al., \textit{this conference}\cite{Smee2024}).

\subsection{Freeform Mirrors} \label{sec:freeform}
\begin{figure}[h]
    \centering
    \gridline{\fig{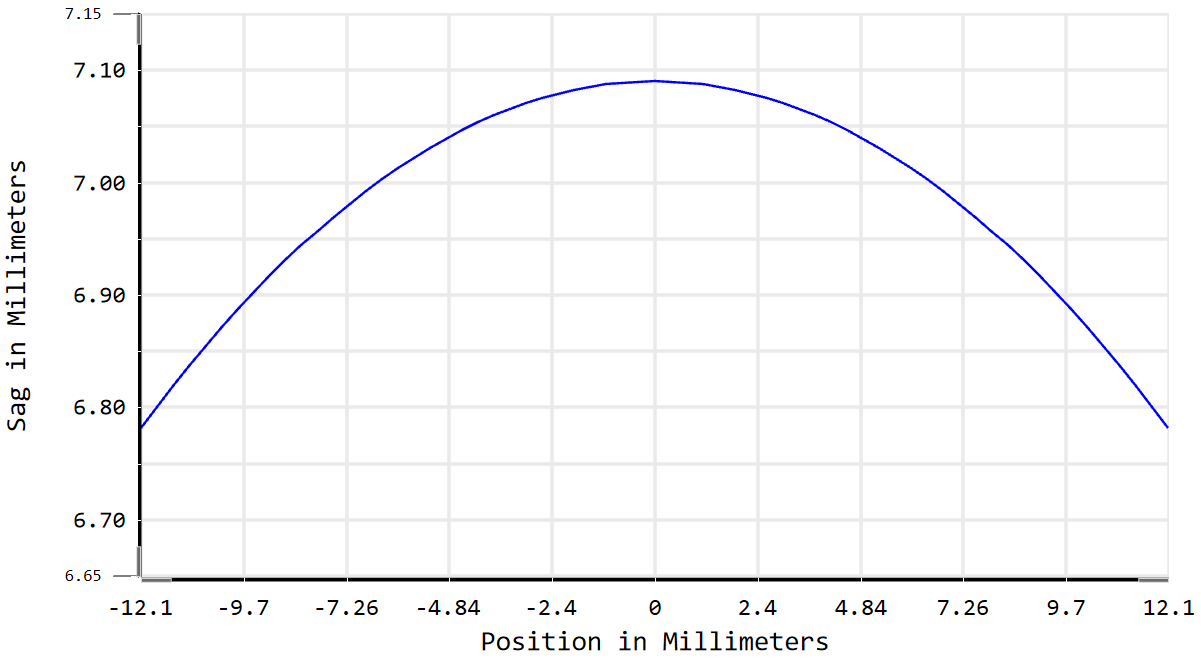}{0.48\textwidth}{(a) Center Slice}
            \fig{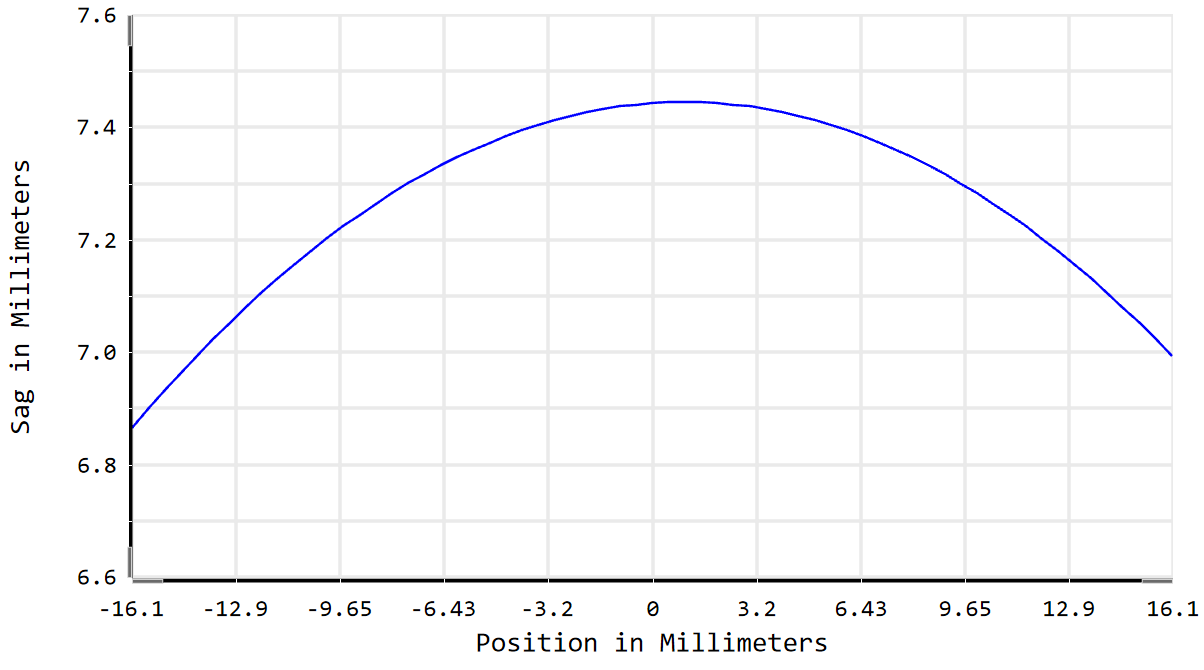}{0.48\textwidth}{(b) Outer Slice}}
    \caption{Cross sections of the pupil mirror sag for the central (a) and outermost (b) slices as a function of position across the mirror diameter. The center of each mirror is at 0mm --- the center of the graphs --- with the full diameter of each mirror captured in these diagnostic plots. Note that due to the direction of the local coordinate system, larger positive sag values indicate a larger offset from the origin --- in other words, the pupil mirrors are all concave. These diagrams illustrate that even though the pupil mirror surfaces are described by a polynomial function, the surface is still smoothly varying. As you can see, there is some asymmetry in the outer slice but the surface is smoothly varying across the full mirror diameter.}
    \label{fig:pupil_sag}
\end{figure}

The pupil mirrors directly follow the slicer and are the most complex element in the system. Each mirror is $\rm \sim25 - 30\,mm$ in diameter, with a freeform shape utilizing the Zemax ``Zernike Phase Sag" surface. This is described by the equation below which matches the ``Even Asphere" surfaces with additional Zernike terms:
\begin{equation}
    z = \frac{c r^2}{1+\sqrt{1-(1+k)c^2r^2}}+\Sigma_{i=1}^8 \alpha_i r^{2i} + \Sigma_{i=1}^N A_iZ_i(\rho, \varphi) \label{eqn:zernike_surface}
\end{equation}
where $r$ is the radial coordinate, $c$ is the curvature, $k$ is the conic constant, $N$ is the number of Zernike terms, $A_i$ are the coefficients on the Zernike terms $Z_i$, $\rho$ is the normalized radial coordinate, and $\varphi$ is the angular ray coordinate. The $\alpha_i r^{2i}$ terms describe even aspheres which are all set to zero in our design. Each pupil mirror is optimized with a unique radius of curvature and a unique freeform shape driven by the Zernike coefficients, $A_i$. For this design we fix all $\rm A_1=A_2=A_3=0$ since these become redundant with the surface tilts and distance from the slicer. Further, we do not extend beyond $\rm A_9$, with the corresponding Zernike terms used described in Table \ref{tab:zernike_terms}. This produces surfaces which are unique and not axially symmetric, but still give smoothly varying surface curvatures in order to be manufacturable. The degree of asymmetry in the mirror curvature varies, with the central slice being the closest to symmetric and the highest degree of asymmetry at the outer slices. Examples of the sag cross sections for pupil mirrors at these two extremes are shown in Figure \ref{fig:pupil_sag}.

\begin{table}[h]
    \centering
    \caption{Zernike terms used in Equation \ref{eqn:zernike_surface} to describe the pupil mirror surfaces.}
    \begin{tabular}{c|c|c}
         $i$ & $Z_i(\rho, \varphi)$  & Name \\
         \hline
         1 & 1 & Piston \\
         2 & $\rho \, cos\varphi$ & Horizontal Tilt\\
         3 & $\rho \, sin\varphi$ & Vertical Tilt\\
         4 & $2\rho^2 - 1$ & Defocus \\
         5 & $\rho^2 \, cos (2\varphi)$ & Vertical Astigmatism \\
         6 & $\rho^2 \, sin (2\varphi)$ & Oblique Astigmatism \\
         7 & $(3\rho^3 - 2\rho) \,  cos\varphi$ & Horizontal Coma \\
         8 & $(3\rho^3 - 2\rho) \,  sin\varphi$ & Vertical Coma \\
         9 & $6\rho^4 - 6\rho^2 +1$ & Primary Spherical\\
         \multicolumn{3}{p{0.4\linewidth}}{NOTE - The coefficients for the first three terms are set to zero in this design.} \\
    \end{tabular}
    \label{tab:zernike_terms}
\end{table}

\begin{figure}
    \centering
    \includegraphics[width=0.95\textwidth]{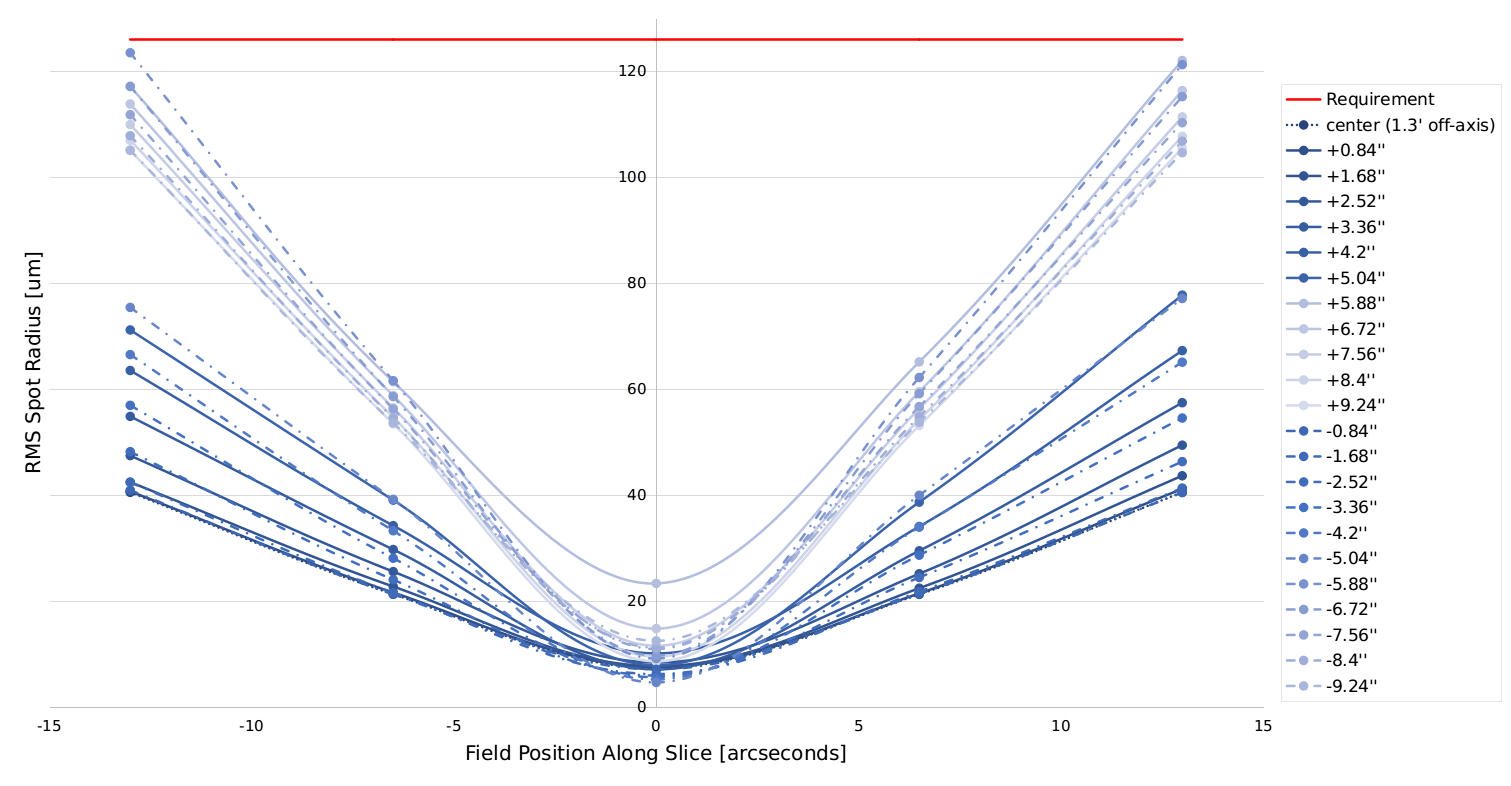}
    \caption{RMS spot radii as a function of position along the slice for all 23 slice positions across the IFU. The horizontal axis shows the position along the slice while each series is labeled by its field position relative to the center slice. Slices offset to the positive direction (further off-axis) are drawn with a solid line while slices offset to the negative direction are drawn with a dashed line. The red line shows the $\rm FWHM<0.6''$ spot size requirement which is equivalent to $\rm126$\um \, RMS radius. All field positions produce spots less than this value.}
    \label{fig:IFU_spots}
\end{figure}

This slicer IFU designs produces spot sizes, shown in Figure \ref{fig:IFU_spots}, with $\rm FWHM<0.6''$ at the focal plane for all slices and field positions, meeting our requirement to produce spots smaller than the median seeing at LCO\cite{Persson1990}. Subpixel dithering\cite{Smee2024} will be used to fully sample the PSF across the slices. Additionally, there will be $\rm >10$ pixels between the spectra of adjacent slices. This is more than sufficient to ensure robust spectral extraction, and is instead limited by mechanical constraints in packaging the IFU between the fixed corrector and CSU surfaces.

The advantage to using these complex optics is an improvement of image quality across the wide field of view, as well as a reduction in overall components. If we were to instead design this IFU with standard spherical pupil mirrors, we would also need to include a set of lens doublets for each individual slice in order to meet our image quality requirements while still maintaining a distance of $\rm 140\,mm$ between the IFU and CSU. This would require not only additional space between slices to mount the doublets, but also additional space between the first fold mirror and the CSU, making the instrument as a whole larger at this location. This would have three important effects: (i) an overall increase in the complexity of the IFU assembly, (ii) a decrease in space available for the telescope guiders at the front of the instrument, and, (iii) a reduction in the FoV possible to include in the IFU mode due to the space constraint of the clear aperture provided by the CSU.

\subsection{Tolerances}
A full tolerance analysis is beyond the scope of this proceedings, but we performed an initial check of the sensitivity to the positions of the mirrors that make up the central and outermost slices to ensure the design strategy is generally feasible. We include in this analysis changes in the tip/tilt and decenter of each mirror in x, and y -- where x and y are perpendicular to the direction of beam propagation -- as well as changes in the distance between each mirror. 

We test default tolerances of $\rm \pm0.2mm$ on each of the decenter and thickness parameters as well as $\rm \pm0.2^\circ$ on tip and tilt, with changes in the RMS spot radius as the evaluation criterion. For the central slice the estimated performance based on the root-sum-square method results in a $<6$\um \, increase in the RMS spot radius for the central slice. 1000 Monte Carlo iterations show that we could expect $<5$\um \, increase 98\% of the time with these tolerances. At the outer slice this increases to a $\sim8$\um \, change. In both extremes this is $<10\%$ of the total spot size requirement with positioning tolerances which seem fully achievable with CNC machining and a precision positioning stage as conceptualized in the following section.

\section{MECHANICAL CONCEPT} \label{sec:mech}
\begin{figure}
    \centering
    \gridline{\fig{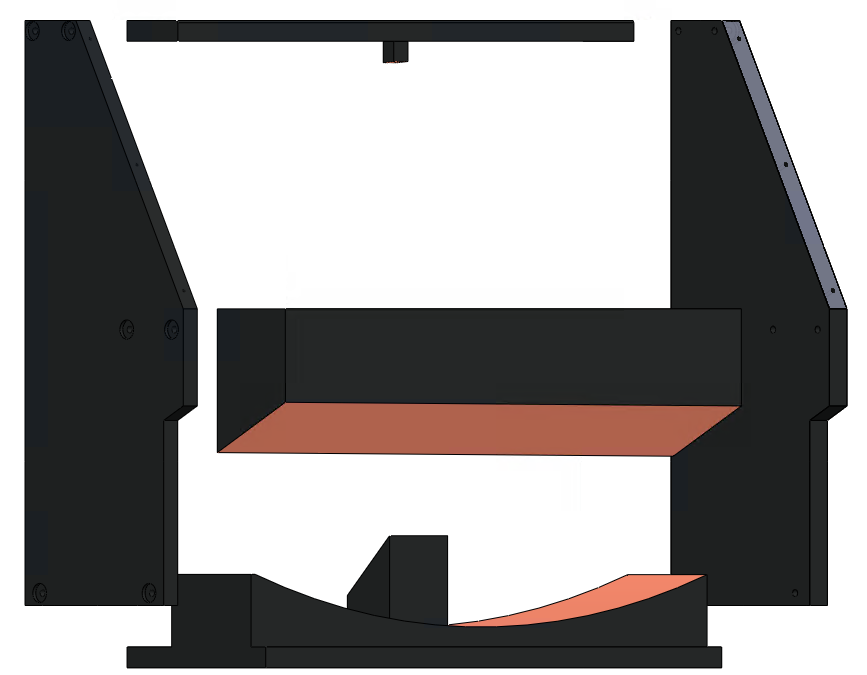}{0.52\textwidth}{(a) Exploded View}
            \fig{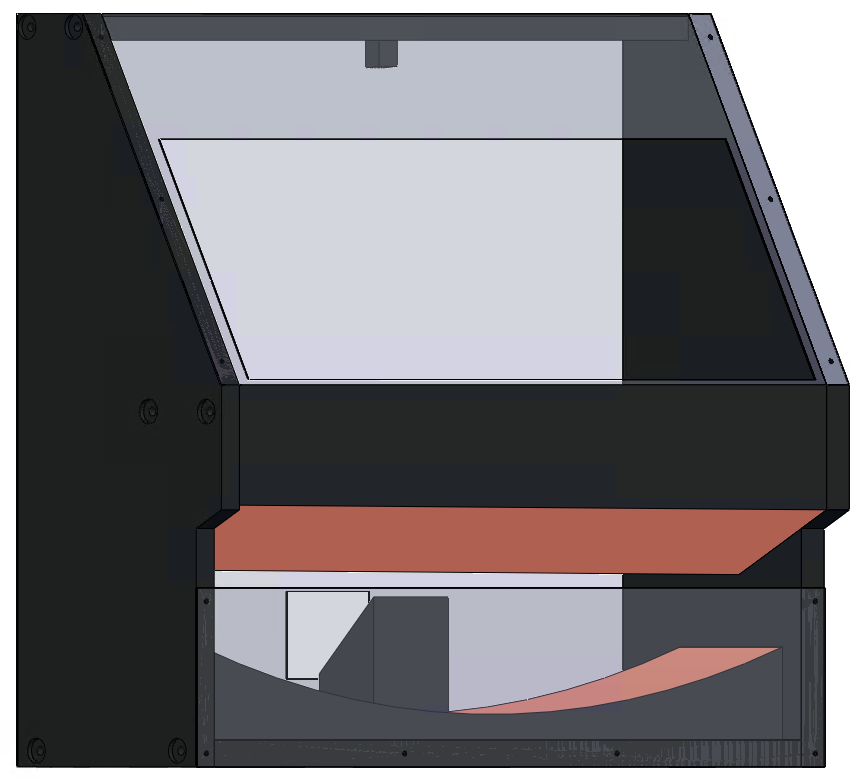}{0.45\textwidth}{(b) Assembled View}}
    \caption{Concept of the machining strategy for the IFU optics and housing. (a) The left side shows an exploded view in order to highlight the individual components. There are three optical pieces: (i) the first fold mirror and the pupil mirrors (bottom) could be machined from a single piece or attached in an intermediate step; (ii) the slicer array (top) would be machined as a monolith; (iii) and lastly, the final fold mirror array (middle). Note that this is a simplified model that does not show the distinct facets that would be present from the individual mirrors in each array. These three components would be attached, aligned, and fixed to each other with the two plates at each side. (b) The right side shows these components assembled with sheet metal baffling included on the front and back sides (shown as transparent here). The large rectangular cut-out is for light to pass through to the CSU while the IFU is in the stored position so that the required range of motion can be reduced.}
    \label{fig:IFU_housing}
\end{figure}

In order to improve and maintain alignment of the system the IFU will be constructed from as few pieces as possible. The slicer array will be fabricated as a monolith and we will aim to do the same for the pupil and fold mirror arrays (elements 3 and 4 respectively). These would then be aligned and fixed to each other with something like the bookend plates shown in Figure \ref{fig:IFU_housing}.

\begin{figure}
    \centering
    \centering
    \gridline{\fig{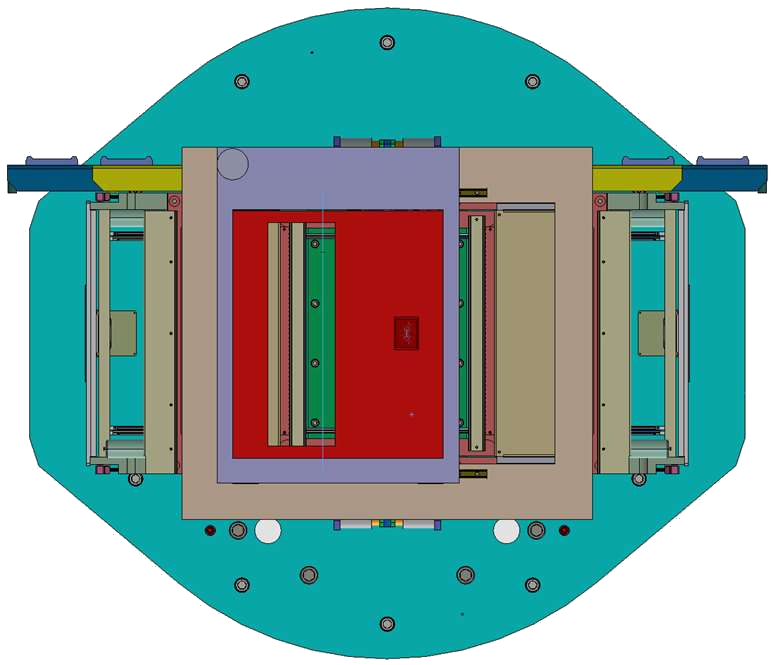}{0.48\textwidth}{(a) Active Position}
            \fig{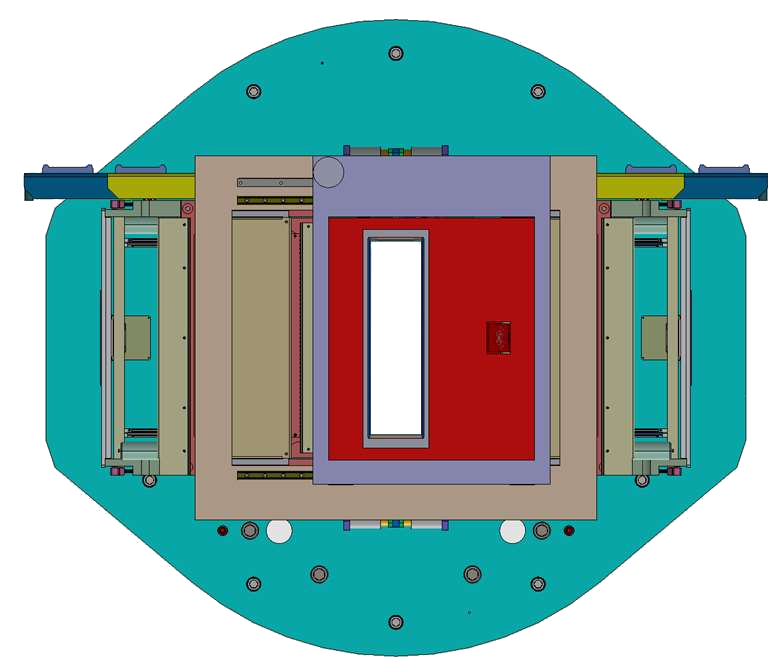}{0.48\textwidth}{(b) Stored Position}}
    \caption{Conceptual design of the IFU mounting stage. The IFU optics concept of Figure \ref{fig:IFU_housing} are highlighted here in red. This is mounted to the gray plate which will have adjustment points to align the IFU with the incoming beam. This plate will in turn mount to the two rails shown above and below the IFU and will be driven by rack and pinion with a cryogenic stepper motor. The active position is shown in (a) with the stored position in (b) highlighting how the light will pass through the window between the slicer and fold mirror arrays in order to reach the CSU when the instrument is used in the MOS mode.}
    \label{fig:IFUstage}
\end{figure}

This entire IFU will be mounted on a linear stage at the front of the CSU mechanism as shown in Figure \ref{fig:IFUstage} in order to shuttle the IFU between the active and stored positions. In this conceptual design we allow for a $50$\um \, tolerance in the spatial position and $\rm 0.5mRad \, (0.03^\circ)$ in tip and tilt. These tolerance allocations leave $\rm\pm150$\um \, decenter and $\rm\pm0.17^\circ$ tip and tilt for the CNC machining of the mirrors and IFU housing. Spring loaded detents that are engaged by default in the active position are base-lined in order to improve the repeatability of positioning. With this scheme the alignment of the optics within the IFU (e.g., slicer to pupil mirrors) will not change as the instrument is switched in and out of the IFU-mode. To simplify the design of this stage and increase it's reliability we will impose an operational constraint such that it can only be moved at a single gravity vector. In order to switch from MOS to IFU mode, the instrument will need to be rotated to this position, the CSU bars fully opened, and the IFU moved into position. At that point the observing target can be acquired and the instrument rotated as needed. We do not expect this to have an appreciable impact on observing efficiency as it is intended that the instrument only be changed between these modes once per night (and we expect most observers to choose only one mode for a given science program).

\section{TESTING}
Under our baseline plan metrology will be performed by the selected vendor to confirm the internal alignment of the IFU optics and that the performance requirements are met. This will be verified at Carnegie Observatories prior to integration with the full MIRMOS instrument.

The linear stage will undergo rigorous lifetime testing with a dummy test mass prior to integration. For a twenty year instrument lifetime we conservatively estimate that the IFU will be moved 8,000 times (20 years $\times$ 200 nights/year $\times$ 2 moves/night = 8,000 moves). We will test that the mechanisms positioning accuracy does not degrade over 3\% of this lifetime or 240 moves. This will be performed in a test cryostat designed to be rotated. The lifetime testing will be performed at a single optimal orientation as we only plan to move the IFU at one gravity vector, but the ability to maintain accurate positioning will be tested at multiple orientations as MIRMOS and the IFU will rotate during use.

\section{SUMMARY}
We have described the preliminary optical design for the MIRMOS IFU as well as a conceptual design for the mechanical stage it will be mounted to. The IFU is a slicer style with 23 slices covering a FoV of $\rm \sim20''\times26''$. The design utilizes freeform surfaces for the pupil mirrors in order to meet our image quality requirements while adhering to the mechanical constraints of the instrument architecture. When built in 2028, this will be the largest FoV IFS operating in the near-IR making it an ideal instrument for a wide range of science cases.

\acknowledgments 
We thank the entire MIRMOS team and our collaborators at CSEM, Teledyne, Universal Cryogenics, and INAF. This material is based on substantial funding from Carnegie Science. This research was funded by the Heising-Simons Foundation through grant 2021-2614. This material is based upon work supported by the National Science Foundation under Grant No. 2206374. We recognize generous support from the Ahmanson Foundation. The Mt. Cuba Astronomical Foundation funded prototype work. M.C. is supported by a Brinson Prize Postdoctoral Fellowship.

\bibliography{mirmos_ifu}
\bibliographystyle{spiebib}

\end{document}